\documentstyle[12pt,amssymb]{article}

\begin{document}

\def\onehalf{{\textstyle \frac12}}
\def\tsty#1#2{{\textstyle\frac{#1}{#2}}}
\def\trans{^{\scriptscriptstyle\top}}
\def\ie{{\it i.e.}}
\def\ssr#1{{\scriptscriptstyle\rm #1}}
\def\vecdos#1#2{\left({\matrix{#1\cr #2\cr}}\right)}
\def\matdos#1#2#3#4{\left({\matrix{#1&#2\cr #3&#4\cr}}\right)}
\def\jour#1#2#3#4{{\sl #1{}} {\bf #2}, #3\ (#4)}
\def\ii{{\rm i}}
\def\lab#1{\label{eq:#1}}
\def\rf#1{(\ref{eq:#1})}
\def\figbox#1#2{\framebox{\hbox to#1cm{\vbox to#2cm{\vfil}\hfil}}}

\newcommand{\be}{\begin{equation}}
\newcommand{\ee}{\end{equation}}
\newcommand{\bea}{\begin{eqnarray}}
\newcommand{\eea}{\end{eqnarray}}

\begin{center}
{\Large\bf Discrete quantum model\\[10pt] of the harmonic oscillator}
\\[1.5cm]
{\sc Natig M.\ Atakishiyev},\footnote{E-mail:
\tt natig@matcuer.unam.mx}\\[7pt]
Instituto de Matem\'aticas\\ Universidad Nacional Aut\'onoma de
M\'exico\\ Av.\ Universidad s/n, Cuernavaca, Morelos 62251,
M\'exico,\\[0.5cm]
{\sc Anatoliy U.\ Klimyk},\footnote{E-mail: \tt aklimyk@bitp.kiev.ua}\\[7pt]
Bogolyubov Institute for Theoretical Physics\\
Metrologichna 14b,  Kiev 03143, Ukraine,\\[0.5cm]
{\sc Kurt Bernardo Wolf},\footnote{E-mail:
\tt bwolf@fis.unam.mx}\\[7pt]
Instituto de Ciencias F\'{\i}sicas\\ Universidad Nacional Aut\'onoma de
M\'exico\\ Av.\ Universidad s/n, Cuernavaca, Morelos 62251, M\'exico.

\bigskip
\bigskip

\end{center}

\begin{abstract}

We construct a new model of the quantum oscillator, whose energy
spectrum is equally-spaced and lower-bound, whereas the spectra
of position and of momentum are a denumerable non-degenerate set
of points in $[-1,1]$ that depends on the deformation parameter
$q\in(0,1)$. We provide its explicit wavefunctions, both in position
and momentum representations, in terms of the discrete $q$-Hermite
polynomials. We build a Hilbert space with a unique measure, where
an analogue of the fractional Fourier transform is defined in order
to govern the time evolution of this discrete oscillator.  In the
limit when $q\to1^-$ one recovers the ordinary quantum harmonic
oscillator.

\end{abstract}


\section{Introduction} \label{sec:one}

Several algebraic constructions have been proposed in the literature
to describe various extensions of the quantum harmonic oscillator.
These constructions are based on various deformations of the standard
oscillator Lie algebra, or different associative algebras. In most of
these models it is difficult to construct a theory for such oscillators,
which is as complete as the well-known treatment of the standard harmonic
oscillator in quantum mechanics. Namely: a canonical complementarity
between position and momentum, explicit forms for the wavefunctions,
and a coherent description of time evolution.

The earliest model, generally called the $q$-oscillator, was proposed
by Macfarlane \cite{Macfarlane} and Biedenharn \cite{Biedenharn} on
the basis of raising and lowering eigenstates of a Hamiltonian with
the $q$-deformed commutator $a^+_q\,a_q\,-\,q\,a_q\,a^+_q$. A theory
of this oscillator has been elaborated (see, {\it e.g.}\
\cite{KulDam}--\cite{AUK}); yet, it has not been clear how to
construct position and momentum operators satisfying the basic
commutation relations with a Hamiltonian to characterize
infinitesimal harmonic motion.  This may be one of the reasons why
this $q$-oscillator has not proven attractive for many physicists.

\smallskip

The postulates we use to define oscillator models are the following
\cite{AAW99,AKW}:

\smallskip

\noindent{\bf 1.}\quad There exists an essentially self-adjoint
{\it position} operator $Q$, whose spectrum ${\cal X}$ is the set
of positions $\{x\}$ of the system.

\smallskip

\noindent{\bf 2.}\quad There exists a self-adjoint and compact
{\it Hamiltonian\/} operator $H$, whose commutator with position
defines the momentum operator $P$,
\be
    [H,Q]=:-\ii\, P,  \lab{def-momentum}
\ee
and corresponds to the first Hamilton equation ($\ii=\sqrt{-1}$);
the commutator of the Hamiltonian with momentum returns the position
operator
\be
    [H,P]=\ii\, Q,    \lab{def-dynamics}
\ee
that corresponds with the second Hamilton equation, and which characterizes
the oscillator dynamics. Equivalent to \rf{def-momentum}--\rf{def-dynamics},
one can propose the Newton-Lie equation as $[H,[H, Q]]= Q$. The set of
momentum values of the system is the spectrum of $P$, which is equal to
that of $Q$ because \rf{def-momentum}--\rf{def-dynamics} generate a rotation
between these two operators (to be written below).

\medskip

\noindent{\bf 3.}\quad The three operators, $Q$, $P$ and $H$, close into
an {\it associative algebra}, \ie, they satisfy the Jacobi identity,
\be
    [P,[H,Q]]+[Q,[P,H]]+[H,[Q,P]]=0.  \lab{Jacobi-identity}
\ee

\medskip

We note that the basic commutator $[Q,P]$ has not been defined.  Due
to \rf{def-momentum}--\rf{def-dynamics}, the only restriction imposed
by the associativity condition \rf{Jacobi-identity}, since the first
two summands are identically zero, is that $[Q,P]$ must commute with
$H$, and thus be constant under the oscillator motion. This indicates
that each distinct choice of the basic commutator $[Q,P]$ will yield
a distinct model for the oscillator.  If the choice is the Heisenberg
commutator $[Q,P]=\hbar{\hat1}$, one has the standard four-generator
oscillator Lie algebra ${\mathcal H}_4= \hbox{span}\,\{H,Q,P,\hat1\}$
of quantum mechanics (containing the Heisenberg algebra ${\mathcal H}_3
=\hbox{span}\,\{Q,P,\hat1\}$). In one previous work \cite{AW97}, with
the purpose of endowing $Q$ with a finite set of position eigenvalues
$x\in\{-j,-j{+}1,\ldots,j\}$ (we write $x|_{-j}^j$), the basic commutator
was taken to be $[Q,P]=\ii\,[H-(j+\onehalf)\hat1]=:\ii\,J_3$, in a matrix
representation of the Lie algebra ${\rm so}(3)={\rm su}(2) = \hbox{span}\,
\{Q,P,J_3\}$ of spin $j$, so the three operators have the same finite
spectrum ${x}|_{-j}^j$. This model, called the ${\rm su}(2)$ oscillator,
has been applied to study the parallel processing of finite signals and
pixellated images \cite{AtPogVicW}.  The general conditions to include
oscillator dynamics in associative algebras were given in \cite{AAW99}.

In a previous paper of the same authors \cite{AKW}, a $q$-algebraic
associative structure was proposed on the basis of the quantum
algebra ${\rm su}_q(2)$, the Hamiltonian having a lower-bound
equally-spaced spectrum. Using a non-standard basis to define
position and momentum operators that allowed analytic expressions,
their spectra was determined to be a finite set of non-equally spaced
points $x_s=\onehalf\sinh (s\kappa)/\sinh\onehalf\kappa$, with
$s|_{-j}^j$ and $q=e^{-\kappa}\in(0,1)$. Also, explicit expressions
were obtained for the wavefunctions in position and momentum, in
terms of the dual $q$-Kravchuk polynomials, related by a fractional
finite Fourier-$q$-Kravchuk matrix transform, and a natural
representation in a {\it sui generis\/} phase space.  The present
paper is a continuation of the research in \cite{AAK} that
constructed quantum oscillators with continuous bounded spectra for
the position and momentum operators.

In this paper we build an oscillator model on the basis of the Fock
representation of a quadratic associative algebra which is a
$q$-deformation of the standard oscillator algebra; we denote this by
${\mathcal DH}_4$, which will be defined in Section \ref{sec:two},
while the physical interpretation of the participant operators that
characterizes this model are set forth in Section \ref{sec:three}.
The spectrum of the Hamiltonian in the algebra is lower-bound and
equally spaced, as in its standard counterpart.  The position spectrum
and wavefunctions, orthonormal under a specific scalar product over
positions, are obtained explicitly in Section \ref{sec:four}, and the
momentum wavefunctions in Section \ref{sec:five}. This model can be
characterized for having a space of positions given by an infinite
non-degenerate point set contained in the interval $[-1,1]$. The
coordinate and momentum realizations of the oscillator are given in
Sections \ref{sec:six} and \ref{sec:seven}; in terms of these we can
write the harmonic oscillator motion in Section \ref{sec:eight}, with
a summation kernel is a new, which is a (denumerable infinite)
relative of the fractional Fourier transform of the standard case.
Concluding remarks are offered in Section \ref{sec:nine}.

We use the notations that are standard in the theory of basic
hypergeometric functions and $q$-orthogonal polynomials (see, for
example, \cite{Gasper-Rahman}), and we assume throughout that $q$
is a fixed real number in $(0,1)$.


\section{The quadratic algebra ${\mathcal D}_q \,{\mathcal H}_4$}
                                        \label{sec:two}

We define the algebra ${\mathcal D}_q{\mathcal H}_4$ as the associative
algebra generated by a vector basis of elements $I_+,\,I_-,\,I_0$,
satisfying the following commutation relations:
\be
    [I_0,I_{\pm}]=\pm I_{\pm},\qquad
    [I_+,I_-] = q^{I_0}-(1+q)\,q^{2I_0}\,.\lab{basic-rel}
\ee
Equivalently, introducing $I_1:=I_++I_-$ and $I_2:=\ii(I_+-I_-)$, we
can characterize this algebra by
\be
        [I_0, I_1] =-\ii I_2, \quad [I_0, I_2] = \ii I_1, \quad
        [I_1,I_2] = -2\ii( q^{I_0}-(1+q)\,q^{2I_0})\,. \lab{altern-basic-rel}
\ee
The first relation in \rf{basic-rel} can be written in the form
\be
        q^{I_0} I_\pm\, q^{-I_0}=q^{\pm 1}I_\pm.   \lab{equiv}
\ee
This relation and the fact that both $q^{I_0}$ and $q^{2I_0}$ appear
in the second relation of \rf{basic-rel} shows that
${\mathcal D}_q{\mathcal H}_4=\hbox{span}\,\{I_1,I_2,q^{I_0},q^{2I_0}\}$
is a quadratic associative algebra. This is a $q$-deformation of the
oscillator algebra ${\mathcal H}_4$ because
$\lim_{q\to1^-}{\mathcal D}_q{\mathcal H}_4={\mathcal H}_4$. Indeed,
in the  limit $\lim_{q\to1^-}$ we obtain from \rf{altern-basic-rel}
the relations
\be
      [I_0, I_1] =-\ii I_2, \quad [I_0, I_2] = \ii I_1, \quad
            [I_1,I_2] = 2\ii,  \lab{IIIcomrel}
\ee
which are equivalent to the defining relations of ${\mathcal H}_4$.

We are interested in the Fock representation of the algebra
${\mathcal D}_q{\mathcal H}_4$; this is an irreducible representation
constructed on a Hilbert space with the orthonormal basis of vectors
$e_n$, $n\in\{0,1,2,\cdots\,\}$ (\ie, $n|_0^\infty$). In this
representation, using the `box' $q$-number $[a]_q:=(1{-}q^a)/(1{-}q)$,
the operators of the algebra act by raising and lowering the number
$n$ of $e_n$,
\bea
    I_+ e_n =\sqrt{q^{n+1}[n{+}1]_q}\,e_{n+1},&&
    I_- e_n =\sqrt{q^{n}[n]_q}\,e_{n-1}, \lab{raise-lower}\\
    I_0 e_n =n\,e_n, & {i.e.},& q^{I_0} e_n =q^n\,e_n,
                                        \lab{weight}
\eea
and the hermiticity conditions $I_+^*=I_-$ and $I_0^*=I_0$ are
satisfied.

In order to have a functional realization of this representation,
we consider the space ${\cal P}$ of all polynomials in one
supplementary variable $y$, and introduce its basis of monomials
\be
    e_n\leftrightarrow e_n (y):= c_n\,y^n , \quad
    c_n = \frac{q^{n(n-1)/4}}{(q;q)_n^{1/2}}, \qquad n|_0^\infty,
                                \lab{e-polynomials}
\ee
where $(a;q)_n:=(1-a)(1-aq)\dots(1-aq^{n-1})$ and $(a;q)_0=1$. Acting
on analytic functions $f(y)\in\cal P$, the Fock representation can be
written in terms of the scale $T_a$ and $q$-difference $D_q$ operators,
\bea
     I_+=\sqrt{\frac{q}{1{-}q}}\,\,y\, T_{q},
        &&  q^{I_0}=T_q,\quad I_- = \,\left[\,q(1-q)\right]^{1/2} D_q;
                    \lab{int-scale-diff}\\
        T_a\,f(y)=f(ay),&& D_q\,f(y)=\frac{f(y)-f(qy)}{1-q}.
                    \lab{scale-diff}
\eea
This realization of the algebra is equivalent to that in \rf{raise-lower}
--\rf{weight}, with the functions $e_n(y)$ playing the role of the basis
elements $e_n$ as eigenfunctions of the weight operator $I_0$.

Let us now introduce a scalar product into the space of polynomials
${\mathcal{P}}$. This scalar product is of a Fisher-type scalar product
and is given by the formula
 $$
\langle f_1(y),f_2(y)\rangle =f_1(\widetilde D_q)f_2^\dag(y)|_{y=0},
 $$
where $\widetilde D_q:=(1-q)T_{q^{-1}}D_q$,  $f_1$ and $f_2$ are
polynomials, and $f_2^\dag$ denotes the polynomial $f_2$, whose
coefficients are replaced by their complex conjugate ones. In this
formula, we have the action of the difference operator upon the
polynomial $f_2^\dag$. Then it is directly verified that
\be
    \langle e_n,\,e_{n'}\rangle = \delta_{n,n'},\qquad
            n,n'|_{\,0}^{\,\infty}.  \lab{Fock-sc-prod}
\ee
Closing the space $\cal P$ with respect to this scalar product we
obtain a Hilbert space that we denote by ${\mathbb H}$. The space
${\mathbb H}$ consists of functions
\be
    f(y)=\sum_{n=0}^\infty b_n e_n(y)=\sum_{n=0}^\infty b_nc_n y^n
        =\sum_{n=0}^\infty a_n y^n,   \lab{fsum}
\ee
where $a_n=b_nc_n$, and $c_n$ are determined by \rf{e-polynomials}.
Since $\langle e_n,\,e_{n'}\rangle = \delta_{n,n'}$ by definition, for
$f_1(y)=\sum_{n=0}^\infty a_n y^n$ and $f_2(y)=\sum_{n=0}^\infty a'_n
y^n$ we have
\be
    \langle f_1,f_2\rangle= \sum_{n=0}^\infty
    \frac{a_na'_n}{|c_n|^2}.    \lab{langler}
\ee
This means that the Hilbert space ${\mathbb H}$ consists of analytic
functions $f_1(y)=\sum_{n=0}^\infty a_n y^n$ such that
\be
    \Vert f\Vert^2 := \sum_{n=0}^\infty \left| \frac{a_n}{c_n}
            \right|^2 < \infty .   \lab{means}
\ee


\section{Assignment of observables to generators}  \label{sec:three}

The {\it discrete oscillator\/} is a class of oscillator models that
depend on the parameter $q\in(0,1)$, and based on the Fock irreducible
representation of the algebra ${\mathcal D}_q {\mathcal H}_4$, where
the physical observables are assigned to the spectra of self-adjoint
generators of the algebra in the following way:
\bea
    \hbox{position:}&& Q:=\sqrt{(1{-}q)/q}\,I_1, \lab{def-pos}\\
    \hbox{momentum:}&& P:=\sqrt{(1{-}q)/q}\,I_2,\lab{def-mom}\\
    \hbox{Hamiltonian:}&& H:=I_0+\onehalf\hat1. \lab{def-Ham}
\eea
Then, due to \rf{altern-basic-rel}, $H$ exhibits the Hamiltonian
oscillator commutation relations \rf{def-momentum}--\rf{def-dynamics}
with $Q$ and $P$, and determines the basic commutator $[Q,P]$,
through
\bea
   &[H,Q]=-\ii\,P,\qquad [H,P]=\ii\,Q,& \lab{disc-osc-1-2}\\[3pt]
   &\displaystyle[Q,P]
   =2\ii(1-q^{-1})\left[q^{H-1/2}-(1+q)\,q^{2H-1}\right]=:\ii\,F(H).&
                                                     \lab{disc-osc-3}
\eea
The operator $F(H)$ defined in \rf{disc-osc-3} commutes with the
Hamiltonian $H$ and is therefore also diagonal in the Fock basis
$\{e_n\}_{n=0}^\infty$ in \rf{e-polynomials},
\be
    F(H)\,e_n  =2(1-q^{-1})\left[ q^{n}-(1+q)\,q^{2n}\right]\,e_n.
\ee
This basis of ${\mathbb H}$ thus consists of eigenfunctions of a
Hamiltonian with equally-spaced eigenvalues,
\be
    H\,e_n=(n+\onehalf)\,e_n,\qquad n|_0^\infty, \lab{H-eigenf-e}
\ee
coinciding with the energy spectrum of the standard quantum harmonic
oscillator.

    From \rf{disc-osc-1-2}, the time evolution of the discrete
oscillator position and momentum operators is produced by
\be
    \exp(\ii\tau H) =e^{\ii\tau/2}\exp(\ii\tau I_0), \lab{expH-tauI}
\ee
and results in the harmonic motion
\be
    \vecdos{Q(\tau)}{P(\tau)}
    = e^{\ii\tau H}\vecdos QP e^{-\,\ii\tau H}
        = \matdos{\phantom{-}\cos\tau}{\sin\tau}{-\sin\tau}{\cos\tau}
            \vecdos QP,   \lab{rota-ps}
\ee
which for $\tau\in[0,2\pi)$ forms a group U(1) of inner automorphisms
of the pair of operators $Q$ and $P$, that we interpret as rotations
of a phase plane around its origin (still to be studied for this model,
see \cite{AKW}). The phase $e^{\ii\tau/2}$ is due to the energy $\onehalf$
of the ground state $e_0$, while $\exp(\ii\tau I_0)$ is the discrete
counterpart of the fractional Fourier transform for this model, to be
seen in Section \ref{sec:eight} below.

The associative algebra ${\mathcal D}_q {\mathcal H}_4$ is a
$q$-deformation of the standard Heis\-en\-berg-Weyl algebra, where
$I_{\pm}$ in \rf{raise-lower} are recognized as the raising and
lowering operators \be
    a_q^+:=I_+=\onehalf\sqrt{\frac{q}{1-q}}(Q-\ii\,P),\quad
    a_q:=I_-=\onehalf\sqrt{\frac{q}{1-q}}(Q+\ii\,P).   \lab{rasin-low}
\ee
From \rf{basic-rel} it then follows that
\be
    [\lim_{q\to1^-}a_q,\lim_{q\to1^-}a_q^+]=\lim_{q\to1^-}
    \left[(1{+}q)\,q^{2I_0}-q^{I_0}\right]=1,            \lab{lim-aa}
\ee
so we recover the standard oscillator. This is the place to emphasize
the difference between the Macfarlane--Biedenharn $q$-oscillator
\cite{Macfarlane}--\cite{AUK}, which is defined in terms of $q$-commutators
$a_q\,a^+_q-q\,a^+_q\,a_q$, and our discrete oscillator, which is formulated
exclusively with ordinary commutators.


\section{Spectrum and eigenfunctions of the posi\-tion
                operator}  \label{sec:four}

A direct calculation using \rf{raise-lower} shows that in the Fock
eigenbasis $\{e_n\}_{n=0}^\infty$ of the Hamiltonian $H$, the position
operator $Q=\sqrt{q^{-1}-1}I_1$ acts as
\be
    Q\,e_n = \sqrt{q^n(1-q^{n+1})}\,e_{n+1}
            + \sqrt{q^{n-1}(1-q^{n})}\,e_{n-1} .
                        \lab{actionQ}
\ee
Since $|q^n(1-q^{n+1})|\le 1$ for $n|_0^\infty$, the norm
$\Vert Q\Vert$ of $Q$ does not exceed 1, and hence it is a
bounded operator whose eigenvalues $\{x\}$ will lie in the real
interval $[-1,1]$.

\subsection{Eigenvectors of position}  

    To find the eigenvectors $\psi_x(y)$ and the spectrum  $\{x\}
={\cal X}$ of the position operator $Q$,
\bea
     Q\,\psi_x(y)&=&x\,\psi_x(y),\qquad x\in{\cal X},\qquad{}
                    \lab{eigenspectrum}
\eea
we represent $\psi_x(y)$ in the form of a linear combination of
the monomials \rf{e-polynomials},
\bea
    \psi_x (y)&=&\sum_{n=0}^\infty p_n(x)\,e_n(y), \lab{analytic}
\eea
where $p_n(x)$ are coefficients depending on the points of the
spectrum $x\in{\cal X}$.

When we substitute the expansion \rf{analytic} into the equation
\rf{eigenspectrum}, we obtain
\be
    \begin{array}{l} \displaystyle
    \sum_{n=0}^\infty \,p_n(x)\,\sqrt{q^n(1-q^{n+1})}\,\,e_{n+1}\,
            + \,p_n(x)\,\sqrt{q^{n-1}(1-q^{n})}\,\,e_{n-1}\\[7pt]
                            {}\displaystyle\qquad\quad
        {}=x \sum_{n=0}^\infty p_n(x)\,e_n. \end{array}
                        \lab{eigenseries}
\ee
From here we find the following three-term recurrence relation for
the coefficients $p_n(x)$ in \rf{analytic},
\be
    x\, p_n(x)= \sqrt{q^n(1{-}q^{n+1})}\, p_{n+1}(x)
            + \sqrt{q^{n-1}(1{-}q^{n})}\, p_{n-1}(x),
                        \lab{recurr-rel-pnx}
\ee
starting with $p_{-1}(x)=0$ and $p_0(x):=1$ setting the common
constant factor.

We see from \rf{recurr-rel-pnx} that the coefficients $p_n(x)$ in
\rf{analytic} are polynomials in $x$ of degree $n$, which can be
evaluated uniquely. To solve the recurrence relation, we make the
substitution
\be
    p_n(x)=(q;q)^{-1/2}_n q^{-n(n-1)/4}  \widetilde p_n(x), \lab{subst}
\ee
which turns \rf{recurr-rel-pnx} into
\be
  x\, \widetilde p_n(x)
    = \widetilde p_{n+1}(x) + q^{n-1}(1-q^{n})\, \widetilde p_{n-1}(x) .
                \lab{recurr-pprima}
\ee
Comparing this with the recurrence relation for the discrete
$q$-Hermite polynomials of type I, given by ${}_2\phi_1$ basic
hypergeometric polynomials \cite[Eq.\ (3.28.3)]{Kk-Swart},
\bea
    h_n(z;q)&:=&q^{n(n-1)/2} {}_2 \phi_1
            (q^{-n},\, z^{-1}; 0;\; q;\,-\,q\,z), \lab{hnzq-basic}\\
    z\,h_n(z;q)&=& h_{n+1}(z;q) +q^{n-1}(1-q^{n})\,h_{n-1}(z;q),
                    \lab{recurr-rel-hnzq}
\eea
we establish that $\widetilde p_n(x)=h_n(x;q)$. We can thus write the
coefficient polynomials in \rf{analytic} as
\be
    p_n(x)=(q;q)_n^{-1/2}q^{-n(n-1)/4} h_n(x;q). \lab{coefposei}
\ee
From \rf{recurr-rel-pnx} follows that these polynomials have definite
parity: $p_n(x)=(-1)^np_n(x)$.

Collecting these results we write the eigenfunctions $\psi_x
(y)$ of the position operator $Q$ as
\bea
    \psi_x (y)&=&\sum_{n=0}^\infty (q;q)_n^{-1/2}q^{-n(n-1)/4}
                            h_n(x;q)\,e_n(y) \lab{psixy1}\\
    &=& \sum_{n=0}^\infty (q;q)^{-1}_n h_n(x;q) \,y^n,\lab{psixy2}\\
    &=& \frac{(y^2;q^2)_\infty}{(xy;q)_\infty},\lab{psixy3}
\eea
where in the last expression we use the symbol $(a;q)_\infty:=\prod_{n=0}
^\infty\,(1-aq^n)$ and the summation formula in \cite[Eq.\ (3.28.11)]{Kk-Swart}.
Because of the convergence of $(y^2;q^2)_\infty$ in \rf{psixy3} for the basis
$\{\psi_x(y)\}_{x\in\cal X}$, we must restrict the domain of definition of
functions $f(y)\in {\mathbb H}$ to the open disk $|y|<1$. Then the condition
$(xy;q)_\infty<1$ is fulfilled automatically since we saw that the eigenvalues
$x\in\cal X$ of $Q$ are contained in the interval $[-1,1]$.


\subsection{The spectrum of position}   

The spectrum of the self-adjoint position operator $Q\sim I_1=I_++I_-$
can be found from the series \rf{psixy1}; from \rf{actionQ} we see that
in the basis $\{e_n(y)\}_{n=0}^\infty$ the operator $Q$ is a self-adjoint
Jacobi tridiagonal matrix of the form
\be
    {\bf Q} =\left( \matrix{
            b_{0}&  a_{0} &0     &0&0&  \cdots   \cr
            a_{0}&  b_1   &a_{1} &0&0&  \cdots   \cr
            0   &   a_1   &b_{2} &a_2&0&  \cdots  \cr
            0   &   0    &a_{2} &b_3&a_3&  \cdots  \cr
        \vdots&\vdots &\vdots &\vdots &\vdots &\ddots  \cr} \right),
                    \qquad\ a_n\ne 0.
\ee
We can now use the theory of these matrices from \cite[Chap.\
VII]{Berezanskii}, (see also \cite{AUK}) to connect their spectra
with the corresponding measures for orthogonal polynomials. In this
vein, we note that in the Fock basis the position eigenfunctions
$\psi_x (y)$ are expanded in terms of the basis elements
$\{e_n(y)\}_0^\infty$ with the {\it polynomial\/} coefficients
$p_n(x)$ in \rf{analytic}, which are given in terms of discrete
$q$-Hermite polynomials of type I in \rf{coefposei}. According to the
results in \cite[Chap.\ VII]{Berezanskii}, these polynomials are then
orthogonal with respect to a spectral measure ${\rm d}\mu(x)$ of the
operator, which is unique up to a constant factor, on a set ${\cal X}
\subset\Re$ that is the simple spectrum of $Q$.

In finding the spectrum of the position operator $Q$, we recall that
the discrete $q$-Hermite polynomials $h_n(x;q)$ obey the orthogonality
relation
\be
     \begin{array}{l}  \displaystyle
     \int_{-1}^1\, (q^2x^2;q^2)_\infty\,
                    h_k(x;q)\,h_m(x;q)\,{\rm d}_qx \\[7pt]
        \displaystyle   {\qquad\quad}
        =\delta_{k,m}\,(1{-}q)(q^2;q^2)_\infty (-1;q)_\infty\,
                    (q;q)_m\, q^{m(m-1)/2} \\[5pt]
            \displaystyle\qquad\quad
        =2\delta_{k,m}\,(1{-}q)(q;q)_\infty (-q;q)^2_\infty\,
                    (q;q)_m\, q^{m(m-1)/2}, \end{array}
                            \lab{q-herm-orthog}
\ee
where $\int_{-1}^1\, f(x)\,{\rm d}_qx$ is  the symbol of the
$q$-integral (see \cite[Eq.\ (3.28.2)]{Kk-Swart}). This
orthogonality relation can be written in the form of a sum
\cite{Gasper-Rahman},
\be
  \begin{array}{l}  \displaystyle
  \sum_{n=0}^\infty (q^{2n+2};q^2)_\infty\,q^n\Big( h_k(q^n;q)
  \,h_m(q^n;q)+ h_k(-q^n;q)\,h_m(-q^n;q)\Big)\\[7pt]  \displaystyle
  {\quad\qquad}=2\delta_{k,m}\,(q;q)_\infty\,(-q;q)^2_\infty\,
  (q;q)_m\,q^{m(m-1)/2}.
  \end{array}                                     \lab{sumform}
\ee
This means that the spectrum ${\cal X}$ of $Q$ is the
simple set of points
\be
        {\cal X} = \{ q^n,-q^n; \ n|_0^\infty\, \},   \lab{SpectrumX}
\ee
and that the corresponding eigenfunctions are
\be
    \psi_{q^n} (y),\quad \psi_{-q^n} (y),\qquad n|_0^\infty,
\ee
given by \rf{psixy1}--\rf{psixy3}. The spectrum of $Q$ is discrete,
which means that the eigenfunctions $\psi_{\pm q^n} (y)$ form a
denumerable orthogonal basis in the Hilbert space ${\mathbb H}$;
we note that ${\cal X}\subset[-1,1]$ has a unique accumulation
point 0 that does not belong to the set.


\subsection{Normalization of the eigenfunctions}  

The eigenfunctions of $Q$ were determined only up to constant
factors, so we proceed to normalize the eigenfunctions
$\{\psi_{\pm q^s}(y)\}_0^\infty$ in their form \rf{psixy2}.
From \rf{psixy1} and the orthogonality of the basis
$\{e_n(y)\}_{n=0}^\infty$ we obtain
\be
\langle \psi_{x} (y), \psi_{x'} (y) \rangle_{\mathbb H}
=\delta_{x,x'} \sum_{n=0}^\infty \frac{q^{-n(n-1)/2}}{(q;q)_n}
\,h_n (x;q)\, h_n(x';q),                     \lab{scprod-psis}
\ee
where $x$ and $x'$ take values in ${\cal X}=\{q^s,\,-q^s;\;
s|_0^\infty\}$. We can calculate this sum as follows: we build
the functions
\bea
    \widetilde h_n (q^s;q)&:=& \sqrt{ \frac{(q^{2s+2};q^2)_\infty\, q^s}
        {2(q;q)_\infty (-q;q)^2_\infty (q;q)_n\, q^{n(n-1)/2}}}\,
                h_n(q^s;q),   \lab{htilde1}\\
    \widetilde h_n (-q^s;q)&:=& \sqrt{ \frac{(q^{2s+2};q^2)_\infty\, q^s}
        {2(q;q)_\infty (-q;q)^2_\infty (q;q)_n\, q^{n(n-1)/2}}}\,
                h_n(-q^s;q).    \lab{htilde2}
\eea

We can see the $\widetilde h_n(\pm q^s;q)$ as the elements $(n,\pm s)$ of
a matrix of numbers (integer $n\ge0$ numbering rows and $s\ge0$
numbering columns), written as
\be
    \vecdos{\{\widetilde h_n (q^s;q)\}_{n,s=0}^\infty}{
        \{\widetilde h_n (-q^s;q)\}_{n,s=0}^\infty}.  \lab{doub-matrix}
\ee
Columns of this matrix are orthonormal due to the orthogonality relation
\rf{sumform} for the discrete $q$-Hermite polynomials $h_k(z;q)$. In the
infinite dimensional case, the orthonormality of columns does not immediately
lead to the orthonormality of rows. But in accordance with the reasoning
of Ref.\ \cite{AKpol}, one can state that rows of this matrix are also
orthogonal, \ie,
\bea
    \sum_{n=0}^\infty \widetilde h_n (q^s;q)\, \widetilde h_n(q^{s'};q)
                &=&\delta_{s,s'},  \lab{orthH1}\\
    \sum_{n=0}^\infty \widetilde h_n (q^s;q)\, \widetilde h_n (-q^{s'};q)
                &=&0,  \lab{orthH2}\\
    \sum_{n=0}^\infty \widetilde h_n (-q^s;q)\, \widetilde h_n(-q^{s'};q)
                &=&\delta_{s,s'}. \lab{orthH3}
\eea
Substituting \rf{htilde1}--\rf{htilde2} into \rf{orthH1}--\rf{orthH3},
we obtain
\be
    \frac{(q^{2s+2};q^2)_\infty\, q^s}{2(q;q)_\infty\, (-q;q)^2_\infty}
        \sum_{n=0}^\infty \frac{h_n(\pm q^s;q)\,h_n(\pm q^{s'};q)
                            }{(q;q)_n\, q^{n(n-1)/2}}
            =\delta_{s,s'}, \lab{orth1}
\ee
where one has to take only the upper or only the lower signs. Returning
to the scalar product in \rf{scprod-psis}, we find
\be
   \langle\psi_{\pm q^s}(y),\psi_{\pm q^{s'}}(y)\rangle_{\mathbb H}
        =\delta_{s,s'}\, \frac{2(q;q)_\infty \,
                (-q;q)^2_\infty}{q^s\,(q^{2s+2};q^2)_\infty}. \lab{orth2}
\ee
We thus arrive at the functions
\be
    \Psi_x(y)\equiv \Psi_{\pm q^s}(y)
        := \sqrt{\frac{(q^{2s+2};q^2)_\infty\,q^s
                }{2(q;q)_\infty (-q;q)^2_\infty}}\,\,
                        \psi_{\pm q^s}(y)  \lab{orth3}
\ee
which are orthonormal under the scalar product \rf{scprod-psis} in
${\mathbb H}$,
\be
    \langle \Psi_{x}(y),\Psi_{x'}(y)\rangle_{\mathbb{H}}
    =\delta_{x,x'},\ \ \ x,x'\in {\cal X}.         \lab{orthofin}
\ee


\section{Spectrum and eigenfunctions of the momentum operator}
\label{sec:five}

The momentum operator $P\sim I_2=\ii(I_+-I_-)$ acts on the basis
$\{e_n(y)\}_0^\infty$ as
\be
    P\, e_n = \ii\Big( \sqrt{q^n(1-q^{n+1})}\, e_{n+1}
                    -\sqrt{q^{n-1}(1-q^{n})}\, e_{n-1} \Big)
                                    \lab{actionP}
\ee
[{\it cf}.\ \rf{actionQ}]. When we change this basis to another
$\{\widetilde e_n\}_0^\infty$ with $\widetilde e_n=\ii^{n} e_n$,
one can see that in the new basis the momentum operator $P$ acts
as a matrix with the same coefficient elements as the position
operator in Section \ref{sec:four} on the former basis. This means
that the spectrum of momentum $P$ coincides with the spectrum of
position $Q$, namely, ${\rm Spec}\, P=\cal X$, where ${\cal X}$
is given in \rf{SpectrumX}. Similarly, the eigenfunctions of momentum
$P$ can be found in the same way as the eigenfunctions of $Q$, by
using the basis $\{\widetilde e_n\}_0^\infty$.

Let $\phi_p (y)$ satisfy $P \phi_p (y)=p\phi_p (y)$, an eigenfunction
of $P$ corresponding to the eigenvalue $p$, with an expansion in the
mode eigenbasis $\{e_n(y)\}_0^\infty$ given by
\be
    \phi_p (y)=\sum_{n=0}^\infty g_n(p)\,e_n(y), \lab{peigen}
\ee
where $g_n(p)$ are coefficients depending on the momentum $p\in{\cal
X}$. Repeating the process of the previous section, one derives a
three-term recurrence relation for the polynomials $g_n(p)$ and
concludes that
 \be
    g_n(p)=\ii^{n} p_n(p)=\frac{\ii^{n} \,h_n(p;q)
                }{(q;q)_n^{1/2}\,q^{n(n-1)/4}},
                        \lab{gn-eigenfmom}
\ee
where $h_n(z;q)$ are the discrete $q$-Hermite polynomials of type I
from Section \ref{sec:four}. Hence, the eigenfunctions of momentum are
\bea
    \phi_p (y)&=& \sum_{n=0}^\infty
        \frac{\ii^{n}\, h_n(p;q)}{(q;q)_n^{1/2}\,q^{n(n-1)/4}}\,
                e_n(y) \lab{phipsum}\\
            &=&\sum_{n=0}^\infty
        \frac{({\rm i}y)^n}{(q;q)_n} h_n(p;q) \lab{phipsumhh}\\
        &=&\frac{(y^2;q)_\infty}{(\ii yp;q)_\infty},\qquad
            p\in {\cal X}=\{ q^s,-q^s;\ s|_0^\infty\}. \lab{phicomp}
\eea
To find the last two expressions we have used the same method as in the
case of eigenfunctions of position in \rf{psixy1}--\rf{psixy3}.

The normalized eigenfunctions of $P$ are
\be
     \Phi_{p}(y)\equiv\Phi_{\pm q^s}(y)
     =\sqrt{\frac{(q^{2s+2};q^2)_\infty\,q^s}{2(q;q)_\infty\,
        (-q;q)^2_\infty}}\,\, \phi_{\pm q^s} (y),
\ee
satisfying $\langle\Phi_x(y),\Phi_{x'}(y)\rangle_{\mathbb H}
=\delta_{x,x'}$, $x,x'\in {\cal X}$.


\section{Coordinate realization of the discrete oscillator}
                                \label{sec:six}

In Section \ref{sec:three} we constructed a realization of the
discrete oscillator on the space of analytic functions in the
supplementary variable $y$ with the assignment \rf{e-polynomials}.
It is natural to look for a realization of the oscillator on
the space of functions in the position coordinate $x\in\cal X$.

Let $L^2({\cal X})$ be the Hilbert space of square-summable
functions over $x\in{\cal X}$ (the set of positions of the
discrete oscillator), with the scalar product
\be
    \begin{array}{rcl}
    \langle f_1,f_2\rangle_{L^2({\cal X})}
        &:=&\displaystyle \frac1{(q^2;q^2)_\infty\,(-1;q)_\infty}
        \sum_{n=0}^\infty (q^{2n+2};q^2)_\infty\,q^n\\[5pt]
        &&{\qquad\qquad}\times\Big(
            f_1(q^n)f_2^*(q^n)+ f_1(-q^n)f_2^*(-q^n)\Big),
                \end{array}         \lab{sumoverqn}
\ee
where ${}^*$ stands for complex conjugation.

Since the discrete $q$-Hermite polynomials are associated with the
determinate moment problem (see, for example, \cite{AUK} for the
description of this association), the set of polynomials
$\{p_n(x)\}_0^\infty$ in \rf{coefposei} constitute a complete set
of orthonormal functions in the Hilbert space $L^2({\cal X})$.

We construct a one-to-one linear isometry $\Omega$ from the Hilbert
space ${\mathbb H}$,  onto the Hilbert space $L^2({\cal X})$, given by
\be
    \Omega: \  {\mathbb H} \ni  e(y)\to f(x)=\langle e(y),\psi_x(y)
    \rangle_{{\mathbb H}} \in  L^2({\cal X}),  \lab{in-sigma}
\ee
where $\psi_x (y)$ are eigenfunctions \rf{psixy3} of $Q$. It follows
from \rf{psixy1} that
 \be
        {\mathbb H}\ni e_n(y)\to  \langle e_n(y),\psi_x (y)
            \rangle_{{\mathbb H}} = p_n(x).   \lab{in-ni}
 \ee
That is, $\Omega$ maps the basis $\{e_n(y)\}$ of ${\mathbb H}$,
which is orthonormal under the scalar product \rf{orth2}, onto
the basis $\{ p_n(x)\}$ of $L^2({\cal X})$, which is orthonormal
under \rf{sumoverqn}; this means that $\Omega$ is a one-to-one
isometry.

In $L^2({\cal X})$, the operator $Q$ acts through multiplication,
\be
                Q \,f(x)=x\,f(x).  \lab{Qfx-xfx}
\ee
Indeed, since $Q\,\psi_x(y)=x\,\psi_x(y)$ for $\Omega\, e(y)=f(x)=
\langle e(y),\psi_x (y) \rangle_{{\mathbb H}}$, we have
\be
    \begin{array}{rcl}
    \Omega: \, Q\,e(y)&\to& Q\,f(x)
        = \langle Q\, e(y),\psi_x (y)\rangle_{{\mathbb H}}\\[3pt]
        &=&\langle e(y),Q\,\psi_x (y)\rangle_{{\mathbb H}}
        =\langle e(y),x\,\psi_x (y)\rangle_{{\mathbb H}} =x\,f(x).
            \end{array}   \lab{map-bases}
\ee

We can find the action of $Q$, $P$, and $H$ on the basis elements
$\{p_n(x)\}_{n=0}^\infty$ of the Hilbert space $L^2({\cal X})$.
According to the recurrence relation \rf{recurr-rel-pnx}, which
follows from the recurrence relation for the discrete $q$-Hermite
polynomials $h_n(z;q)$, we have for the position operator $Q$ that
\be
    Q\,p_n(x) = \sqrt{q^n(1-q^{n+1})} \,p_{n+1}(x) +
                \sqrt{q^{n-1}(1-q^{n})} \,p_{n-1}(x).
                            \lab{Qacton-p}
\ee

It follows from formulas (3.28.7) and (3.28.8) in \cite{Kk-Swart} that
the momentum operator $P$ acts on the Hilbert space $L^2({\cal X})$
through
\be
    P=-\ii(1-q)\,q^{H-1/2}\Big( D_q
            +\frac1{q^2\,(q^2x^2;q^2)_\infty}D_{q^{-1}}
                            (q^2x^2;q^2)_\infty \Big),
                        \lab{diff-form-P}
\ee
where $(q^2x^2;q^2)_\infty$ is the multiplier in the orthogonality
measure in the scalar product \rf{sumoverqn}. In particular, $P$
acts on the basis functions $\{p_n(x)\}_{n=0}^\infty$ as
\be
    P\,p_n(x) =\ii \sqrt{q^n(1-q^{n+1})} \,p_{n+1}(x)
                -\ii \sqrt{q^{n-1}(1-q^{n})} \,p_{n-1}(x).
                    \lab{Pacton-p}
\ee

Finally, the Hamiltonian $H$ acts on the basis polynomials $p_n(x)$
of the Hilbert space $L^2({\cal X})$ as
\be
        H\,p_n(x)=(n+\onehalf)\,p_n(x). \lab{Hact}
\ee
Indeed, according to \rf{H-eigenf-e} and \rf{in-ni} we have
\be
    \begin{array}{rcl}
        H\,p_n(x)&=& \langle He_n(y),\psi_x (y)\rangle_{{\mathbb H}}\\[5pt]
            &=&(n+\onehalf) \langle e_n(y),\psi_x (y)\rangle_{{\mathbb H}}
                    =(n+\onehalf)\, p_n(x).
                    \end{array}     \lab{Hactop}
\ee

It is interesting to see the lowest discrete oscillator modes
$p_n(x)$ in \rf{coefposei} and \rf{Hact}, both as continuous
functions of $x\in(0,1]$ and their values at the orthogonality set
$\cal X$, for various values of $q$. While the continuous functions
exhibit strong oscillations (increasing with $n$ and $1/q$), their
values on ${\cal X}(q)$ remain well bounded. As $q\to1^-$, the
corresponding points in $\cal X$ densify, evincing the resemblance
of the discrete oscillator with the standard oscillator
wavefunctions. This form of convergence should be studied further,
since a change of scale appears necessary as well as a discrete
measure that becomes a continuous Riemann integral in the limit. One
analogue for this limit appears in  \cite[Fig.\ 4]{ANVW}, where the
Meixner functions that describe the discrete model converge to the
Laguerre-Gauss modes of a radial oscillator; in that case though,
the limit is from hyperboloids to the cone in the three-dimensional
space of the Lie algebra ${\rm su}(1,1)$, and is not a
$q$-deformation.


\section{Momentum realization of the discrete oscillator}
                                    \label{sec:seven}

Consider the Hilbert space $L^2({{\cal P}})$ of square-integrable
functions $f(p)$ in the momentum coordinate $p$ in the oscillator
with the same scalar product as in \rf{sumoverqn}, where ${\cal P}
= {\cal X}$ is the spectrum of the momentum operator $P$, coinciding
with the spectrum of $Q$. The coefficient polynomials $g_n(x)$ in
formula \rf{gn-eigenfmom} for the eigenfunctions of momentum,
$\phi_p(y)$ in \rf{peigen}, constitute an orthonormal basis in
$L^2({{\cal P}})$.

To formalize this consideration, we construct, as in the previous
Section, a one-to-one linear isometry $\widetilde \Omega$ from the
Hilbert space ${\mathbb H}$ onto the Hilbert space $L^2({{\cal P}})$,
given by
\be
    \widetilde \Omega:  \ {\mathbb H} \ni  e(y)\to f(p)
        :=\langle e(y),\phi_p(y)\rangle_{{\mathbb H}}
                            \in  L^2({{\cal P}}),
            \lab{omega-tilde-map}
\ee
where $\phi_p (y)$ are the eigenfunctions of momentum $P$ in
\rf{phipsum}--\rf{phicomp}. [Compare with \rf{in-sigma} requiring
the position eigenfunctions $\psi_x(y)$ in \rf{psixy1}--\rf{psixy3}.]
From here it is evident that
\be
    {\mathbb H}\ni e_n(y)\to
        \langle e_n(y),\phi_p (y) \rangle_{{\mathbb H}} =g_n(p),
                \lab{map=H-to_l2}
\ee
that is, $\widetilde \Omega$ is a one-to-one isometry and maps the
orthonormal basis $\{e_n(y)\}_0^\infty\in{\mathbb H}$ onto the
orthonormal basis $\{g_n(p)\}_0^\infty\in L^2({{\cal P}})$.

The momentum operator $P$ acts on $L^2({{\cal P}})$ as a
multiplication operator on all functions of $p$,
\be
            P \,g(p)=p\,g(p).  \lab{Pactmom}
\ee
The action of $Q$, $P$, and $H$ on the basis of polynomials
$g_n(p)$ can be found in the form of the recurrence
relations
\bea
    Q\,g_n(p)  &=& \sqrt{q^n(1-q^{n+1})} \,g_{n+1}(p)
                + \sqrt{q^{n-1}(1-q^{n})} \,g_{n-1}(p),
                            \lab{Qongp}\\
    P\,g_n(p)  &=&\ii \sqrt{q^n(1-q^{n+1})} \,g_{n+1}(p)
                -\ii \sqrt{q^{n-1}(1-q^{n})} \,g_{n-1}(p),
                            \lab{Pongp}\\
    H\, g_n(p) &=&(n+\onehalf)\, g_n(p).
                            \lab{Hongp}
\eea


\section{Harmonic evolution in position space}  \label{sec:eight}

According to \rf{expH-tauI} and \rf{rota-ps}, the action of the operator
$\exp(\ii\tau H)$ is the time evolution of the discrete oscillator.  On
the basis \rf{e-polynomials} of functions $\{e(y)\}_0^\infty$ that is
orthonormal with respect to the scalar product \rf{Fock-sc-prod}, this
action is
\be
    e^{\ii\tau H}\, e_n(y)= e^{\ii\tau/2}\, e^{\ii n\tau}\,e_n(y)
            = e^{\ii(n+1/2)\tau}\, e_n(y).
\ee
The operator $\exp(\ii\tau H)$ also acts on the Hilbert space
$L^2({\cal X})$, which is characterized by the scalar product
\rf{sumoverqn}. Now consider the isometry between these two spaces,
\be
    {\mathbb H}\ni e(y) \to f(x):=
        \langle e(y),\psi_x(y)\rangle_{\mathbb H} \in L^2({\cal X}),
                            \lab{isometry-hl}
\ee
that maps functions $e(y)$ onto functions $f(x)$ of the discrete
position coordinate $x\in\{-\,q^n,q^n\}_0^\infty={\cal X}$. Then
to $\exp ({\rm i}\tau H)e(y)\in {\mathbb H}$ there corresponds a
function $\exp ({\rm i}\tau H)f(\pm q^s)$ of $x=\pm\,q^s$,
\bea
    e^{\ii\tau H}f(\pm\,q^s)&=&  \langle e^{\ii\tau H} e(y),
        \psi_{\pm\,q^s}(y)\rangle_{\mathbb H}
        = \langle e(y),e^{-\ii\tau H}\psi_{\pm\,q^s}(y)\rangle_{{\mathbb H}}
                        \lab{Fou-1}\\
 &=&\sum_{n=0}^\infty \langle e(y),e_n\rangle_{{\mathbb H}}\langle e_n,
        e^{-\ii\tau H} \psi_{\pm\,q^s} (y)\rangle_{{\mathbb H}}
                        \lab{Fou-2}\\
 &=&\sum_{n=0}^\infty \langle e(y),e_n\rangle_{{\mathbb H}}
        \langle e^{\ii\tau H} e_n ,\psi_{\pm\,q^s} (y)\rangle_{{\mathbb H}}
                        \lab{Fou-3}\\
 &=& \sum_{n=0}^\infty \sum_{m=0}^\infty \Big(
        \langle e(y), \Psi_{q^m} (y) \rangle_{{\mathbb H}}
        \langle \Psi_{q^m} (y), e_n\rangle_{{\mathbb H}}
                        \nonumber\\
    &&{\quad\phantom{\sum\sum}} +
        \langle e(y), \Psi_{-\,q^m} (y) \rangle_{{\mathbb H}}
        \langle \Psi_{-\,q^m} (y), e_n\rangle_{{\mathbb H}}\Big)
                        \lab{Fou-4}\\
     &&{\qquad\phantom{\sum\sum}}\times
            e^{\ii\tau (n{+}1/2)}\langle e_n,
        \psi_{\pm\,q^s}(y)\rangle_{{\mathbb H}}         \nonumber\\
 &=& \sum_{m=0}^\infty\Big(K^\tau(\pm\,q^s,q^m)f(q^m)+
         K^\tau(\pm\,q^s,-q^m)f(-q^m)\Big);
                        \lab{Fou-5}
\eea
where, according to \rf{psixy1},
\bea
    K^\tau(\pm\,q^s,\pm\,q^m)&=& c_s\sum_{n=0}^\infty
        \langle \psi_{\pm q^m}(y), e_n\rangle_{{\mathbb H}}\,
            e^{\ii\tau (n+1/2)}
        \langle  e_n, \psi_{\pm\,q^s}(y)\rangle_{{\mathbb H}}
                        \lab{Fou-K}\\
  &=& c_s e^{\ii\tau/2} \sum _{n=0}^\infty
        \frac{q^{-n(n-1)/2}}{(q;q)_n}\,
            h_n(\pm\,q^m;q)\,e^{\ii n\tau}\,h_n(\pm\,q^s;q)
                        \lab{Fou-K2}
\eea
and
\bea
    c_s&:=& \frac{q^s(q^{2s+2};q^2)_\infty}{2(q;q)_\infty
            (-q;q)^2_\infty} .         \lab{Fou-cs}
\eea

Being elements of $L^2({\cal X})$, the functions $f(x)$, $x\in\cal X$,
enter into the scalar product \rf{sumoverqn} as a sum with the weight
function $(x^2q^2;q^2)_\infty$. Because we intend to interpret these
as wavefunctions of a quantum oscillator, and display them for comparison
with their standard shapes, we should absorb this weight into the functions,
\be
        F(x)=\sqrt{(x^2q^2;q^2)_\infty}\,f(x),  \lab{absorb}
\ee
so that their scalar product acquires, from \rf{sumoverqn}, the
`standard' form,
$$
\langle F_1,F_2\rangle_{\cal X}:=\frac{1}{(q^2;q^2)_{\infty}\,
(-1;q)_{\infty}}\,\sum_{x\in{\cal X}}\,|x|\,F_1(x)\,F_2(x)^*\,,$$
\be
x\equiv x(\pm,n)=\pm\,q^n\in{\cal X},\quad n|_0^\infty\,. \lab{sumoverx}
\ee
The matrix elements of operators will be correspondingly rescaled by
\rf{absorb}.

The oscillator evolution \rf{Fou-1}--\rf{Fou-5} can be used to define
the fractional discrete Fourier transform on the functions
$F(x)$, $x\in{\cal X}$, rescaled as in \rf{sumoverx}. We note that
the fractional Fourier integral transform for angle $\tau$ differs
from the standard harmonic oscillator evolution by a phase $e^{\ii\tau/2}$
that is due to the ground energy $\onehalf$ in the oscillator, so the
fractional Fourier transform is
\be
    \Phi(\tau):=e^{-\ii\tau/2}\exp(\ii\tau H).  \lab{Fou-oscc}
\ee
Its action on the rescaled discrete wavefunctions will have the form
\bea
    \Phi(\tau)\,F(x)&=&\sum_{x'\in{\cal X}} \Phi(x,x';\tau)\,F(x'),
                                \lab{FT-1}\\
    \Phi(x,x';\tau)&=& e^{-\ii\tau/2}\sqrt{\frac{(x^2q^2;q^2)_\infty}
    {(x^{\prime2}q^2;q^2)_\infty}}\,K^\tau(x,x'),  \lab{FT-2}
\eea
where $x,\,x'\in{\cal X}$, and $K^\tau(x,x')$ is given by
\rf{Fou-K}--\rf{Fou-K2} for $x=\pm\,q^s$ and $x'=\pm\,q^m$.
Unfortunately, the bilinear generating function \rf{Fou-K2} of
discrete $q$-Hermite polynomials of type I could not be summed
to a closed form.


\section{Concluding remarks}  \label{sec:nine}

We constructed a model of the harmonic oscillator that can be realized
on bases of coordinate and momentum Hilbert spaces, and its energy
modes expressed in terms of discrete $q$-Hermite polynomials of type I.
The spectrum of the Hamiltonian coincides with that of the standard
harmonic oscillator in quantum mechanics, while the position and
momentum operators in this model have discrete, denumerably infinite
spectra that depend on the extension parameter $q$ contained in the
interval $[-1,1]$.

Contrary to other models that use discrete $q$-Hermite polynomials
\cite{Lorek,Hinterding,Alvarez}, our models (the present one and that
in Ref.\ \cite{AKW}) fulfill the basic Hamilton equations in the form
$[H,Q] = -\ii P$ and $[H,P]=\ii Q$, with {\it standard commutators \/}
--- and not $q$-commutators \cite{Macfarlane,Biedenharn}. Because of
this important circumstance, the time evolution of the model is a Lie
group, which but for a phase is that of fractional Fourier transforms
associated with this model \cite{Montreal}. This discrete oscillator
is a new and non-trivial deformation of the standard quantum harmonic
oscillator; it allows the extension of other standard concepts of phase
space, such as coherent states \cite{AKW}, that will be examined elsewhere.

We believe that the discrete oscillator model can appropriately
describe discrete quantum systems on bounded point lattices, and also
contribute significantly to the general theory of special functions.

\section*{Acknowledgements}

This research was supported by the SEP-CONACYT (M\'exico) project
IN\-102\-603 {\it \'Optica Matem\'atica}, and  Grant 14.01/016 of
the State Foundation of Fundamental Research of Ukraine.



\begin{thebibliography}{99}

\bibitem{Macfarlane} A.J.\ Macfarlane, On $q$-analogues of the
quantum harmonic oscillator and the quantum group $SU(2)_q$,
\jour{J.\ Phys.\ A: Math. Gen.}{22}{4581--4588}{1989}.

\bibitem{Biedenharn} L.C.\ Biedenharn, The quantum group $SU_q(2)$
    and a $q$-analogue of the boson operators, \jour{J.\ Phys.\ A:
    Math. Gen.}{22}{L873--L878}{1989}.

\bibitem{KulDam} P. P.\ Kulish and E. V. \ Damaskinsky,  On the $q$
    oscillator and the quantum algebra $su_q(1,1)$, \jour{J.\ Phys.
    \ A: Math. Gen.}{23}{L415--L419}{1990}.

\bibitem{AtakSus} N.M.\ Atakishiyev and S.K.\ Suslov, Difference
    analogs of the harmonic oscillator, \jour{Teor.\ Math.\ Phys.}
    {85}{1055--1062}{1991}.

\bibitem{Burban} I.M.\ Burban and A.U. \ Klimyk, On spectral properties
    of $q$-oscillator operators, \jour{Lett.\ Math.\ Phys.}{29}{13--18}
    {1999}.

\bibitem{AUK} A.U.\ Klimyk, On position and momentum operators in the
    $q$-oscillator, \jour{J.\ Phys.\ A:\ Math.\ Gen.}{38}{4447--4458}{2005}.

\bibitem{AAW99} M.\ Ar\i k, N.M.\ Atakishiyev, and K.B.\ Wolf, Quantum
    algebraic structures compatible with the harmonic oscillator Newton
    equation, \jour{J.\ Phys.\ A: \ Math.\ Gen.}{32}{L371--L376}{1999}.

\bibitem{AKW} N.M.\ Atakishiyev, A.U.\ Klimyk, and K.B.\ Wolf, Finite
    $q$-oscillator, \jour{J.\ Phys.\ A:\ Math.\ Gen.}{37}{5569--5587}{2004}.

\bibitem{AW97} N.M.\ Atakishiyev and K.B.\ Wolf, Fractional
    Fourier-Kravchuk transform, \jour{J.\ Opt.\ Soc. Am.\
    A}{14}{1467--1477}{1997}.

\bibitem{AtPogVicW} N.M.\ Atakishiyev, G.\ Pogosyan, L.E.\ Vicent, and
    K.B.\ Wolf, Finite two-dimensional oscillator.\ I: The Cartesian
    model, \jour{J.\ Phys.\ A: \ Math.\ Gen.}{34}{9381--9398}{2004};
    \ II: The radial model, \jour{J.\ Phys.\ A: \ Math.\ Gen.}{34}
    {9399--9415}{2004}.

\bibitem{AAK} M.N.\ Atakishiyev, N.M.\ Atakishiyev, and A.U.\ Klimyk,
    On $su_q(1,1)$-models of quantum oscillator, \jour{J.\ Math.\ Phys.}
    {47}{093502}{2006}.

\bibitem{Gasper-Rahman} G.\ Gasper and M.\ Rahman, {\it Basic
    Hypergeometric Functions\/} (Cambridge University Press,  2004).

\bibitem{Kk-Swart} R.\ Koekoek and R.F.\ Swarttouw, {\it The
    Askey-Scheme of Hypergeometric Orthogonal Polynomials and Its
    $q$-Analogue} (Delft University of Technology, Report 98--17,
    1998 [available from {\tt ftp.tudelft.nl}]).

\bibitem{Berezanskii} Yu.M.\ Berezanski\u{\i}, {\it Expansions in
    Eigenfunctions of Selfadjoint Operators\/} (Providence, R.I.,
    American Mathematical Society, 1969).

\bibitem{AKpol} N.M.\ Atakishiyev and A.U.\ Klimyk, On $q$-orthogonal
    polynomials, dual to little and big $q$-Jacobi polynomials,
    \jour{J.\ Math.\ Anal.\ Appl.}{294}{246--257}{2004}.

\bibitem{ANVW} N.M.\ Atakishiyev, Sh.M.\ Nagiyev, L.E.\ Vicent, and
    K.B.\ Wolf, Covariant discretization of axis-symmetric linear
    optical systems, \jour{J.\ Opt.\ Soc.\ Am.\ A}{17}{2301--2314}{2000}.

\bibitem{Lorek} A.\ Lorek, A.\ Ruffing, and J.\ Wess, A $q$-deformation
    of the harmonic oscillator, \jour{Z.\ Phys.\ C}{74}{369--377}{1997}.

\bibitem{Hinterding} R.\ Hinterding and J.\ Wess, $q$-deformed Hermite
    polynomials in $q$-quantum mechanics, \jour{Eur.\ Phys.\ J.\ C}{6}
    {183--186}{1999}.

\bibitem{Alvarez} R.\ \'Alvarez-Nodarse, M.K.\ Atakishiyeva, and N.M.\
    Atakishiyev, On a $q$-extension of the linear harmonic oscillator
    with the continuous orthogonality property on $\mathbb R$, \jour
    {Czech.\ J.\ Phys.}{55}{1315--1320}{2005}.

\bibitem{Montreal} K.B.\ Wolf, Discrete and finite fractional Fourier
    transforms. In: {\it Procee\-dings of the Workshop on Group Theory
    and Numerical Methods\/} (Universit\'e de Montr\'eal, 26--31 May,
    2003), {\sl CRM Proceedings and Lecture Series\/} Vol.\ {\bf 39},
    267--276 (2004).

\end{thebibliography}
\end{document}